# Enhanced initial protein adsorption on engineered nanostructured cubic zirconia

R. F. Sabirianov[1*], A. Rubinstein[1*], F. Namavar[2]

[1] *Department of Physics, University of Nebraska at Omaha, Omaha, NE 68182*
[2] *Department of Orthopaedic Surgery and Rehabilitation, University of Nebraska Medical Center, Omaha, NE 68198*
*Corresponding authors: rsabirianov@mail.unomaha.edu and arubinstein@mail.unomaha.edu

Motivated by experimentally-observed biocompatibility enhancement of nanoengineered cubic zirconia ($ZrO_2$) coatings to mesenchymal stromal cells, we have carried out computational analysis of the initial immobilization of one known structural fragment of the adhesive protein (fibronectin) on the corresponding surface. We constructed an atomistic model of the $ZrO_2$ nano-hillock of 3-fold symmetry based on Atom Force Microscopy (AFM) and Transmission Electron Microscopy (TEM) images. First principle, quantum mechanical calculations show a substantial variation of electrostatic potential at the hillock due to the presence of surface features such as edges and vertexes. Using an implemented Monte Carlo (MC) simulated annealing method, we found the orientation of the immobilized protein on the $ZrO_2$ surface (both atomically flat and nanostructured) and the contribution of the amino acid residue from the protein sequence to the adsorption energy. Accounting for the variation of the dielectric permittivity at the protein-implant interface, we used a model distance-dependent dielectric function to describe the inter-atom electrostatic (Coulomb) interactions in the adsorption potential. We found that the initial immobilization of the rigid protein fragment on the nanostructured pyramidal $ZrO_2$ surface is achieved with a magnitude of adsorption energy larger than that of the protein on the smooth (atomically flat) surface. The strong attractive electrostatic interactions are a major contributing factor in the enhanced adsorption at nanostructured surface. In the case of adsorption on the flat, uncharged surface this factor is negligible. We show that the best electrostatic and steric fit of the protein to the inorganic surface corresponds to a minimum of the adsorption energy determined by the non-covalent interactions.

## Introduction

Protein adsorption on a solid artificial surface is a fundamental phenomenon that determines the biological response of a living organism entering any implant material. The adsorbed proteins play the role of mediator in interactions between cells and implants[1,2]. The biocompatibility and bio-integration of implanted devices with the affected tissue depends very strongly on the initial immobilization and possible structural changes of proteins on the biomaterial surface. During the initial immobilization, the protein orientation must be properly controlled, so that the corresponding cell receptors can interact with the functional sites of the immobilized proteins. The problem of controlling protein adsorption (in particular, the initial protein immobilization on solid inorganic surfaces) remains a challenge due to the limited number of studies completed in this field so far[1,3,4]. The modeling of the adsorption process is very complicated and should include an adequate description of the solid surface properties (roughness, defects, crystallographic texture, electric potential variation across the surfaces, and charge distribution) and the solvent dielectric properties at the protein-implant interface. Despite the importance of this problem, and in contrast to the abundance of literature on the modeling of protein-protein complexes, similar modeling has not been performed on protein-implant complexes. This is due to a lack of experimental knowledge about the atomic coordinates of any of such complexes. Subsequently, this shortage of information complicates the testing of prospective models. So far, the majority of modeling studies of protein (as well as peptide) adsorption is performed on smooth (i.e. atomically flat) surfaces using detailed atomistic approaches such as Molecular Dynamics[5-8], Monte Carlo[9-13] and Brownian Dynamics[14,15] methods, as well as hybrid approaches including minimization techniques[4,16-18]. While protein adsorption on artificially-engineered nanostructured solid surfaces (both theoretically and computationally) has not been adequately explored, recent experimental studies by several groups show that nanostructured solid surfaces may enhance cell growth and proliferation, as well as affect differentiation[1,19-22]. For example, Namavar *et al.* found that the growth and proliferation of bone marrow stromal cells on nanostructured $ZrO_2$ (fake diamond) coatings is enhanced compared with conventional orthopaedic metallic and ceramic smooth surfaces[20]. Remarkably, it was also shown that these coatings[23,24] are superior even to bulk hydroxyapatite (HA) in terms of cell adhesion and growth. Combined with its superior mechanical and chemical stability, these findings make $ZrO_2$ a very promising implant material[20,23].

The ability of the implant surface to adsorb proteins determines its aptitude to support cell adhesion, spreading, and biocompatibility[25]. In the case of the solid surface, the protein adsorption is *unspecific*, since specific lock-and-key mechanisms (realized in protein-protein complex formation) are absent[26]. Active studies during recent years in the field of biomolecular materials show that the non-covalent (long-range electrostatic and short-range van der Waals) interactions between the protein and artificial surface are major factors in the protein adsorption.[1,4,6,14,20] While modeling of the van der Waals (vdW) interactions is based on the well-known approaches[27], the evaluation of electrostatic interactions (EI) is far from straightforward, particularly in the presence of an aqueous solvent[28,29]. The dielectric properties of the solvent near protein and inorganic surfaces are poorly determined. It is known that these properties are very different from the bulk properties of solvent.[28,29] The presence of surface features such as edges, steps and hillocks (particularly at nano-scale) further complicates the description of the EI.

The adsorption of protein on solid implants can be considered analogous to the formation of protein-protein complexes. The analysis of many protein associations has shown that the oppositely-charged and polar residues located in the vicinity of binding sites tend to form inter-protein complementary electrostatic contacts[30-37]. It was found, both experimentally and theoretically, that the favorable EI between the two associating proteins could be an essential factor for this binding process[30-51]. Specifically, according to the aforementioned theoretical works[30,33,37,47,48], the existence of favorable electrostatic binding thoroughly explains the experimentally-measured rate constants for protein association. In the case of protein complex formation both steric (shape fit) and electrostatic complementarity of the interacting proteins contributes to the lock-and-key specific recognition mechanism[20,30,41,42]. This is the major difference from the unspecific protein adsorption on the inorganic surface. However, we hypothesize that these complementary sites responsible for the protein recognition may be mimicked by engineering the nanostructured inorganic surfaces. These sites (nanostructured surface regions) are expected to provide sufficiently strong binding of protein to the surface and provide the necessary protein orientation. These ideas can explain the recent observation that hut-shaped nano-hillocks of tantalum oxide with sharper edges adsorb fibronectin (FN) better than surfaces of other shapes[22]. The control of protein binding (including its orientation) is reminiscent of the inorganic surface modification by chemically attaching active protein binding fragments (peptides) to the surface. However, granting that native proteins will attach to a pure nanostructured surface and promote the cell adhesion, this chemical treatment is not necessary.

Cell attachment and spreading *in vitro* is generally mediated by several adhesive proteins, among which FN plays an important role for cell adhesion. Fibronectin is a large extracellular matrix protein, consisting of globular domains (connected by short flexible segments), one of which contains the cell receptor-binding amino acid sequence motif Arg-Gly-Asp (RGD). This protein is known to pay a crucial role in adsorption-dependent cellular activities including attachment, proliferation and differentiation[1,52,22]. The orientation of FN on the surface, as well as how strongly it is bound to the surface, are thought to be important parameters for the ability to promote cellular binding. It was found that the solid implant surface [1,22,53] treated with FN creates cell adhesion and the density of coverage can be used to control this adhesion. Because it is one of the main adhesive proteins, FN is extensively studied experimentally and the crystallographic structure of its fragments is available. These factors make FN a very attractive object for theoretical modeling of protein adsorption on artificial surfaces.

In the present work, we model the initial immobilization of a rigid protein fragment of FN (with known 3D structure) containing the 13-14 type 3 repeats (13FN3-14FN3)[54] on the engineered nanostructured solid $ZrO_2$ surface[20,23,24]. We chose this fragment because it contains the large positively charged cluster of amino acid residues associated with heparin binding site (HBS) that should be attracted to the negatively charged sites of $ZrO_2$ surface. We investigate the influence of the surface topology (roughness) on the physical properties of nanostructured surfaces, and discuss its relationship to protein adsorption. Using the quantum mechanical calculations, we show that there is a spatial variation of electrostatic potential on the artificial surface due to the presence of surface crystallographic features (vertices, edges, steps, and vacancies). Our results indicate that the FN fragment adheres to the $ZrO_2$ surface in such a way that distribution of the positively charged amino acid residues in the protein is compatible and complementary to the charge distribution of our engineered $ZrO_2$ surface.

Our calculation suggests that the nanostructured surface possesses areas of high charge density, while the smooth areas have a lower variation of surface charge density. Based on our calculations, the electrostatic potential difference of our nanostructured pyramidal hillock is up to half of a volt, similar to the surface of protein[55]. Using MC simulations method we evaluated the adsorption energy determined by the total non-covalent, electrostatic and vdW, interactions between the protein and the surface atoms. We show that immobilization of protein on the nanostructured pyramidal $ZrO_2$ surface can be achieved with an absolute adsorption energy larger than that for typical protein adsorption on a smooth (flat) surface. Our calculation shows that the location and orientation of proteins on the surface is determined by the favorable EI between effective negatively charged zirconia nanostructures and positively charged residue cluster of the 13FN3 fragment responsible for the heparin binding site (HBS).

**Model and Simulation Method**

An atomic 3D structure model of the protein is available in the Protein Data Bank (PDB, www.pdb.org), while the model of the artificial solid surface should be determined from the experimental data. The latter can vary substantially even for the same material depending on the preparation conditions. The 3D AFM and 2D TEM images of the nanostructured $ZrO_2$ surface (see Figure 5 in Ref.[23]) are typical for films fabricated by ion beam assisted deposition (IBAD)[23,24]. The pyramidal hillocks with 3-fold symmetry are observed by AFM and TEM images on a nanostructured $ZrO_2$ surface[23,24].

*Atomistic model of a nanocrystal of cubic zirconia.* Based on the AFM and TEM images, as well as typical surface terminations of $ZrO_2$, we have built a simplified model of a hillock. The atomic structure of the $ZrO_2$ hillock in the shape of the three-fold pyramid is given in Figure 1. We cut the hillock out of periodic lattice using cleavage planes chosen to mimic experimentally-observed surface terminations based on AFM data. This single pyramid has four (111) faces representing the flat surfaces. It contains edges created by the intersection of two (111) planes and two types of vertices. An oxygen (O) atom at the top shows sharper vertex, while three vertices at the lower face show a more rounded example (with the O atom removed).

*First principle electronic structure calculations of the nanostructured surface.* We performed first principle, quantum mechanical calculations[56] for the $ZrO_2$ pyramidal nano-hillock with three-fold symmetry. We obtain electronic structure, electrostatic potential and charge transfer in the $ZrO_2$ nano-hillock. We use a $Zr_{20}O_{42}$ tetrahedron cluster terminated by (111) planes as shown in Figure 1. Calculations were performed using density functional method in general gradient approximation (implemented in commercially available VASP package *cms.mpi.univie.ac.at/vasp/*) by using Γ-point in reciprocal space and $E_{cutoff}$=350eV.

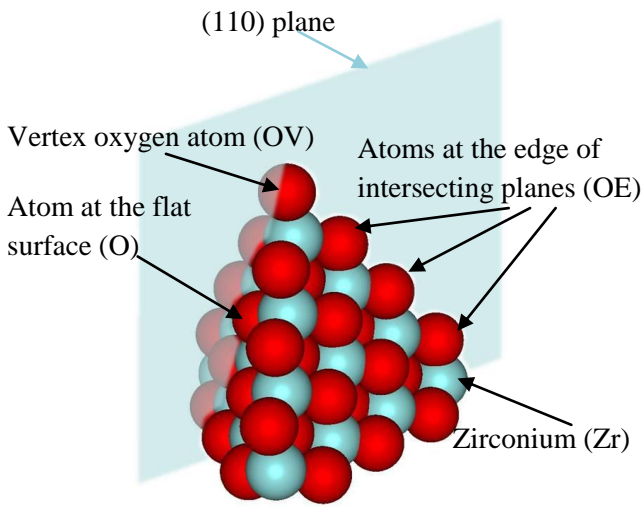

Figure 1. Crystal structure of $ZrO_2$. The crystallite is formed by cleavage of Zr crystal by (111) planes. The (110) cross-section plane containing representative features of nanostructured surfaces such as edges and vertexes is shown (and will be used to show electrostatic potential in Figure 4).

*Atomistic model for cubic zirconia slab.* In order to compare the initial immobilization of the rigid 13FN3-14FN4 fragment on the smooth and nanostructured surfaces, we built two models by cutting out a finite slab from the periodic bulk crystal:

(1) The smooth (flat) surface is modeled by a slab of $ZrO_2$ terminated by (111) planes as shown in Figure 2A. This model contains 7290 Zr and O atoms. The thickness of the slab is 2 unit cells or ~15Å. The slab is stoichiometric and does not possess a net charge.

(2) Based on AFM images, we built an atomistic model of a nanostructured surface of $ZrO_2$ textured along (111) direction. The surface consists of three pyramids cleaved by (111) planes and one complementary "smaller" pyramid formed by (001)-plane cleavage from periodic lattice as shown in Figure 2B. The vertex of the latter pyramid is cut off by (111) plane. The distance between the vertices of the larger pyramids is about 3 nm, while the distances between the vertices of the larger pyramids and the top of the middle pyramid is 2 nm. The above atomistic model contains 2901 Zr and O atoms and is presented in coordinate file by CHARMM format. The total charge of the atomic model is of -2.2e.

*Protein model.* The 13FN3-14FN3 structured domain of human FN, found to be responsible for heparin binding[54], is considered in atomistic detail and is constrained to be rigid in our simulations. We consider a model of this domain from the larger 12FN3-14FN3 domain with known 3D structure (**1fnh.pdb**), taken from the PDB by truncating the linker residues Leu91 and Glu92 between the type 3 repeat 12 and 13 of FN (12FN3 and 13FN3, respectively). The binding of this fragment to heparin is due to the EI between positively charged basic amino acid residues (Arg/Lys) grouped in a space cluster and the negatively charged groups of heparin (sulfate and carboxylate groups)[54].

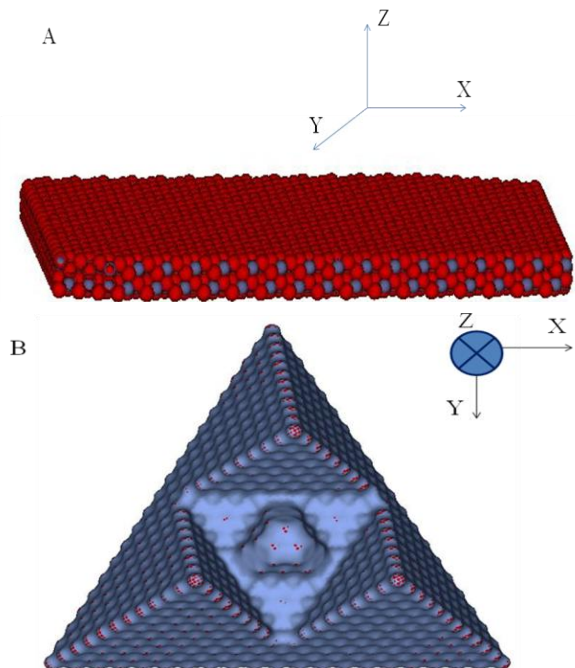

Figure 2. (**A**) Smooth (atomically flat) surface modeled by a slab of $ZrO_2$ terminated by (111) planes. The Zr and O atoms are shown by gray and red, respectively. (**B**) Atomistic model of the typical pyramidal fragment of the nanostructured $ZrO_2$ surface designed by IBAD technique. Red indicates the negatively charged O atoms of the flat surfaces, vertexes and edges of the pyramids.

The first part of the considered domain, 13FN3, has amino acid sequence N93-T181. The second part of the domain, 14FN3, has amino acid sequence I183-T271. These are connected by the linker amino acid residue A182. It was previously shown that 13FN3 fragment provides the dominant HBS containing a large cluster of the positively charged residues (Arg 98, 99, 101, 115, 146 and Lys117)[54, 57]. Additionally, the 13FN3-14FN3 domain is a particularly rigid unit in comparison with other FN domains[53], and is considered as rigid in our MC simulations. We hypothesize that electrostatic interactions between this positively charged fragment and the negatively charged $ZrO_2$ solid surface would enhance protein adsorption. Simple estimations indicate that the protein 13FN3-14FN3 domain has a positive net charge of +9e at neutral pH of solution, that is a result of balancing 24 positively (Arg/Lys) and 15 negatively (Glu/Asp) charged residues located in this subunit.

*All-atom model for adsorption potential.* The model adsorption potential $U_{total}$ consists of screened Coulombic and vdW (approximated by a Lennard-Jones 6-12 potential) potential energy functions $U_{el}$ and $U_{vdW}$ for protein-surface inter-atom interactions

$$U_{total} = \sum_{i,j} U_{el}(r_{ij}) + \sum_{i,j} U_{vdW}(r_{ij}) = \sum_{i,j} q_i q_j / \varepsilon(r_{ij}) r_{ij} + \sum_{i,j} \varepsilon_{ij} \{(A_{ij}/r_{ij})^{12} - 2(A_{ij}/r_{ij})^6\}, \quad (1)$$

where $r_{ij}$ is the distance between atom $i$ of a protein and atom $j$ of an inorganic artificial surface; the summation is over all $i$ and $j$ atoms; $q_i$ and $q_j$ are the correspondent partial charges of the atoms located at the protein and the surface, respectively; $\varepsilon(r_{ij})$ is an effective dielectric permittivity at the interface that is considered in the next subsection. The vdW parameters $\varepsilon_{ij}$ for the cross interactions between $i$ and $j$ atoms are obtained from those of the pure components using the Lorentz-Berthelot mixing rules: $\varepsilon_{ij} = (\varepsilon_i \varepsilon_j)^{1/2}$, where $\varepsilon_i$ and $\varepsilon_j$ are the corresponding minimum energy of the potential curves typical for the atom type[58]. The parameters $A_{ij} = r_{min,i}/2 + r_{min,j}/2$, where $r_{min,i}$ and $r_{min,j}$ are the distances that correspond to the half-widths of the above potential curves[58]. We use CHARMM force field parameters[59] including the partial charges of amino acid residue atoms ($q_i$) and the corresponding vdW constants ($\varepsilon_i$ and $r_{min,i}$) for protein atoms. The partial atom charges ($q_j$), depending on the location on the $ZrO_2$ surface, are calculated the first time from the electronic structure calculations of the nanostructured solid surface (see "Results"). The vdW parameters for atoms of $ZrO_2$ ($\varepsilon_j$ and $r_{min,j}$) are fitted from the published data[60,61] (Table 1).

Table 1. Lennard-Jones parameters used in force field for atoms of $ZrO_2$.

|    | $\varepsilon_i$ (kcal/mol) | $r_{min,i}/2$ (Å) |
|----|----|----|
| Zr | -0.003891 | 2.0 |
| O  | -0.15 | 1.815 |

We use atom-atom pair approximation $\sim(A_{ij}/r_{ij})^6$ at longer distances in (Eq.1) for calculation of vdW interactions. It is known that the vdW interaction energy between the atom and the semi-infinite dielectric surface depends as $\sim 1/D^3$ (D is an atom-surface distance)[27]. However, in the case of the nanostructured surface, the calculation of vdW interactions cannot use approximation of planar (or spherical) surfaces at shorter distances when the sizes of the surface hillock features (such as edges and vertexes) are of the same order as the distance from the atom to the surface. Because vDW interactions are due to induced atomic dipole polarization, it will be substantially screened at the larger distances in aqueous polar solvent (it is true both for atomically flat and nanostructured surfaces)[27]. Thus, the use of finite dimensions of the surface slabs, i.e. finite interacting region (Figure 2), is fully justified. It is effectively similar to use a cutoff distance of these interactions traditionally used in Molecular Dynamics simulations[4,62,63].

*Model for the effective dielectric permittivity function $\varepsilon(r_{ij})$.*
Despite a large number of studies to date, accurate estimations of the EI energy between two bio-molecules in a solvent continue to be computationally very demanding, particularly in a simulation of protein-protein association[28,29,64]. The same problem exists in the case of EI between a protein (or any bio-molecule) and a solid surface of artificial substrate during the adsorption process[1].

We use the distance-dependent dielectric function $\varepsilon(r_{ij})$ analogous to the one of the effective dielectric function obtained for a cross-media pair-wise electrostatic interaction (CPEI) energy between two point charges, when one charge is located in a solvent and the other one in a dielectric medium (see Eq.19 in Ref.[28]):

$$\varepsilon(r_{ij}) = \varepsilon_{Ld} / [1 + (\varepsilon_{Ld}/\varepsilon_{Sd} - 1)\exp(-r_{ij}/\lambda)] \quad (2).$$

This function varies at the length-scale $\sim\lambda$ from the dielectric constant $\sim\varepsilon_{Sd}$ at the small distances ($r_{ij} < \lambda$) to the value of $\sim\varepsilon_{Ld}$ at the long distances ($r_{ij} > \lambda$). The dielectric constant $\varepsilon_{Sd} = 15$ is considered as an average value $\sim(\varepsilon_{ZrO2} + \varepsilon_p)/2$ determined for $ZrO_2$ ($\varepsilon_{ZrO2} = 25$)[65] and protein ($\varepsilon_p = 4$)[66]. We consider the effective dielectric constant $\varepsilon_{Ld} = 50$ that corresponds to an average value $\sim(\varepsilon_{ZrO2} + \varepsilon_{H2O})/2$, where the dielectric constant of the aqueous solvent $\varepsilon_{H2O} \approx 80$. The considered approximation for the dielectric permittivity (2) corresponds to the asymptotic solution, including the classical expressions for the CPEI energy (3) in the case of two uniform dielectrics[67], $ZrO_2$ and protein, for rather small ($r_{ij} \ll \lambda$) as well as $ZrO_2$ and the solvent for rather long ($r_{ij} \gg \lambda$) inter-atom distances.

$$U_{el}(r_{ij}) \approx \frac{q_i q_j}{r_{ij}} \begin{cases} 2/(\varepsilon_{ZrO_2} + \varepsilon_p), & r_{ij} \ll \lambda \\ 2/(\varepsilon_{ZrO_2} + \varepsilon_{H_2O}), & r_{ij} \gg \lambda \end{cases} \quad (3).$$

The simplified approximation (2)-(3) reflects the fact that when protein is in direct contact with the artificial surface, at small distances between interacting charges, the electric field lines (EFL) concentrate in the local surface regions of the protein–surface interface. In this case, the values of $\varepsilon_{ZrO2}$ and $\varepsilon_p$ are major factors determining an electrostatic energy (3) when $r_{ij} < \lambda$. The use of dielectric function (Eq. 2) is justified because charged amino acid residues in proteins tend to be exposed at the protein surface[68]; and the major contribution to the electrostatic field on the inorganic surface, which is induced by the protein, is determined by the protein charges in proximity to the surface, i.e. $r_{ij} < \lambda$. The larger the interatomic distance, the larger number of EFL across the solvent, resulting in a larger value of the effective dielectric function $\varepsilon(r_{ij})$. When the protein is not in direct contact with the artificial surface, or when the protein ionogenic groups are far from the surface, the EFL between surface charges of interest (in a protein and an artificial surface) cross over the solvent and the CPEIs experience an additional screening due to the high solvent permittivity. In this case, the value of the effective dielectric constant is determined mainly by $\varepsilon_{ZrO2}$ and $\varepsilon_{H2O}$ at the long inter-atomic distances. We use $\lambda \sim 10$Å because it is a typical length-scale of the spatial variation of the cross-media effective dielectric function at the interface dielectric-solvent[28,29]. This parameter determines to what extent the long-range, inter-atomic EI are included in consideration, and, the value of $\lambda$ is analogous to the cutoff length ~10Å used for truncation of the pair-wise EI energy in a free energy simulation of the solute-solvent systems[4,62,63].

*Simulations of protein adsorption.* In the present work, we implemented the MC simulated annealing method[69] using a Metropolis algorithm[58] to find the optimal immobilization of the rigid protein body on the nanostructured surface that corresponds to the minimum of total energy determined by Eq.1. For each fixed temperature in the MC simulated annealing procedure, random changes are made to the current protein location and orientation relative to the surface. The energy of a new state is then compared to one of its predecessors. The new state is accepted if the Boltzmann factor $\exp(-\Delta U_{total}/k_B T)$ is larger than the random number in the interval [0..1]. This procedure is performed for each temperature cycle, spanning the configuration space of the system in search of a minimum of corresponding adsorption energy.

Gradual cooling from a rather high initial temperature of the system is carried out to avoid trapping in the local minima. A typical MC simulation produced an initial annealing temperature range of 500K and 4000K with a temperature reduction by 50K/cycle and 500K/cycle, respectively, with 5000 steps (or configurations of the system) in each cycle till zero temperature is reached. As a result, we have about 50,000 steps of the simulation per each program run. The in-house program was written in FORTRAN90.

The initial position of the protein relative to the modeled $ZrO_2$ slab surface is determined in the following manner. The origin point with fixed axes X, Y and Z, during MC simulations is placed at the geometric center of the typical pyramidal surface fragment (Figure 2B). The normal vector of the $ZrO_2$ (111) crystalline plane is oriented along the z-axis (Figure 2B). Based on this (global) coordinate system, the local coordinate system of the protein rigid body (axes X', Y' and Z') is defined by the origin point placed at the geometrical center of the protein, when the X, Y and X', Y' planes are parallel to each other, while axes Z and Z' coincide. The axis X' is oriented along both structured segments (13FN3 and 14FN4) of the considered 13FN3-14FN3 protein domain with a minimum of the moment of inertia of the protein relative to the center of mass. The geometric centers for both the protein and the $ZrO_2$ slab are separated at the given initial distance $R_Z$ (about 40 to 50 Å) along the z-axis. During MC simulations, translational and rotational motions of the protein relative to the surface on each MC step are performed in the conventional way[58].

**Results and Discussion**

*Electronic properties of the nanocrystalline $ZrO_2$ surface.* Using first principle quantum mechanical calculations[56] for the nanocrystallite $ZrO_2$ in the shape of a pyramid with three-fold symmetry, we obtain the electronic structure, electrostatic potential and charge transfer of $ZrO_2$. The charge density of the above $Zr_{20}O_{42}$ hillock is shown in Figure 3. Thus, the distinguished surface features are captured in one figure. The excess of the O atom on the top and the edges provide effective negative charge on the pyramid. The isosurface of the charge is shown in gray. The charge density of the flat surface has a lower variation of charge density near the surface (seen as semicircles in the isosurface at the edge). Additionally, the top of the pyramid has four O atoms (red) and only one Zr atom (blue) giving effective negative charge at the vertex.

The shape of the isosurface at the vertex is more localized near the atom, suggesting larger charge density on the atom, while charge density near atoms on the flat (111) plane shows the formation of bond with Zr atom (stronger covalent bonding).

We calculate local atomic charges using Bader analysis by separating the space of the system into basins using zero-flux surfaces. Bader analysis for $Zr_{20}O_{42}$ shows that the charge acquired by the O is about 1.5-1.7e, while each Zr atom looses 2.2-3.5e. Additionally, we perform Voronoi tessellation of the charge density and calculate the charge in the Voronoi polyhedrons. During this analysis, the space is separated by splitting the distance between the O and Zr atoms in half. Thus, all the points which are closer to O contribute to the charge on O atoms, while points which are closer to Zr contribute to the charge on Zr. The results of these two methods are qualitatively the same, but the Voronoi charges are slightly smaller.

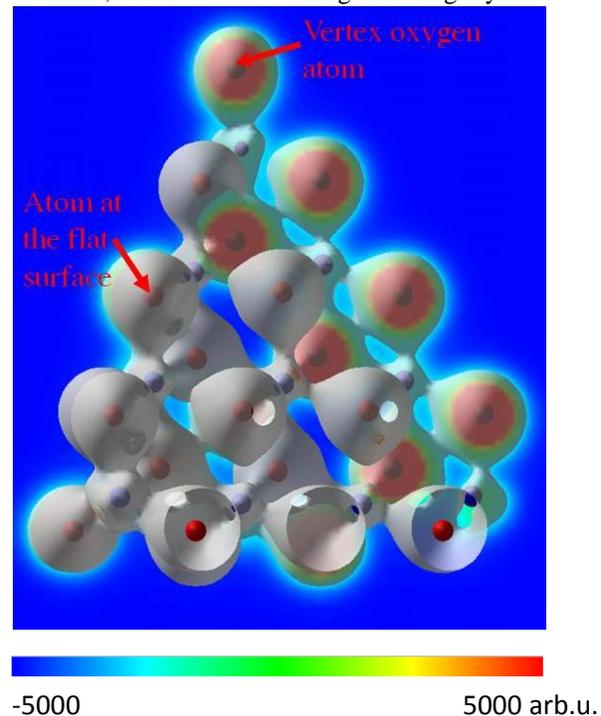

-5000                                              5000 arb.u.

Figure 3. The charge density of Zr nanocrystallite. Red spheres show O atoms (negatively charged) and blue spheres show Zr atoms (positively charged). The isosurface is shown in gray.

We will use Voronoi data for further discussions. The O acquires 0.75÷1.2e, while Zr loses 2.34÷2.5e. In the inner region O receives 1.2e and Zr donates 2.4e. Near the surface the charge distribution shifts compared to the inner regions because of the O termination, resulting in bonds between O and Zr being directed inwards to the particle. Even larger changes are observed at the edge and vertex of O sites. For example, the vertex site has only one Zr neighbor and it acquires 0.75e from this neighbor. Oxygen in the inner region gets about 0.3e from each of its nearest Zr (coordinated by 4 Zr atoms). Thus, the charge density in the nanoparticle is quite different than that of the flat surfaces, creating substantially different electric potential variation. The effective charges of atoms located on the $ZrO_2$ pyramidal surface vary and their estimated values are given in Table 2, together with the effective charge of the smooth (plane) (111) surface.

Table 2. The charges typical for the model nanocrystallite of $ZrO_2$

| Atom | Charge (ξe) |
|---|---|
| Zr | +2.4e |
| O | -1.2e |
| OV | -0.75e |
| OE | -1.12e |
| OB | -1.01e |

*Denotes*: Zr- zirconium atom, O- oxygen atoms in the inner region, OV – oxygen pyramid vertex atom, OE – oxygen pyramid edge atom and OB – oxygen atoms located on the smooth (111) plane surface of the $ZrO_2$ substrate.

The electrostatic potential distribution of the $Zr_{20}O_{42}$ hillock is shown in Figure 4. We use the (110) cross-section of the hillock because it passes through special features: the edges on the right side, the vertices, and the flat face on the left side. The potentials are truncated out at -5eV to reveal the detailed features near the surface. The electrostatic potential near the flat surface, and near the vertices and edges, is substantially different. The common practice in visualization of the potential of proteins is to map them on the solvent accessible surface. This surface is built by rolling a ball of particular size (1.4 Å) on the surface of the protein created by its vdW radii. Thus, assuming that the O has a vdW radius of 1.78Å, we can see that there is substantial difference in the electrostatic potential on the solvent accessible surface near vertices and flat regions. There is also electrostatic potential dependence on the atomic coordination of surface features. The vertices with Zr closer to the surface have lower value of the potential. Thus, we can conclude that surface sharper features like edges and vertices create enhanced variation of electrostatic potential on the nanostructured surface compared to the flat surface.

Thus, *ab-initio* quantum mechanical calculations of the model nanocrystallite $ZrO_2$, clearly indicate that the spatial electric potential variation across our designed insulator surfaces is comparable to the variation electrostatic potential of the proteins[55]. In the next section we discuss our results of initial physical protein adsorption on the smooth and nanostructured $ZrO_2$ surface.

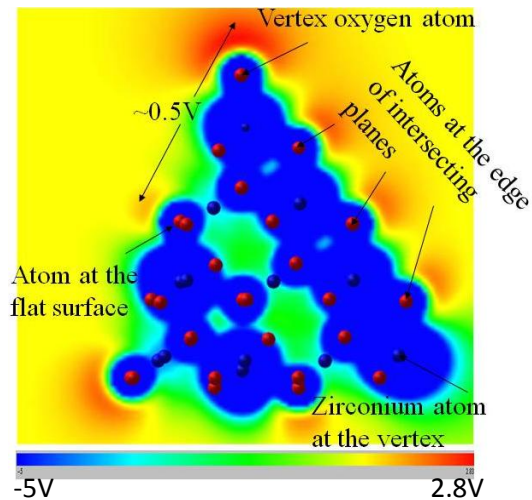

Figure 4. Electric potential of $Zr_{20}O_{42}$ cluster on the cross-section (110)-plane (See Fig. 1). The larger value of potential is observed near the sites with low coordination. The potential near the flat surfaces has somewhat lower value. The voltage between two points of solvent accessible surface of the O atom on the flat surface and the one at the vertex is ~0.5V.

***Protein adsorption on the $ZrO_2$ surface.*** We perform MC simulations to study initial physical adsorption of the 13FN3-14FN3 domain of human FN on the flat surface of the $ZrO_2$ slab (Figure 2A). These calculations are carried out using simulated annealing starting from protein being initially positioned on top of the geometric center of $ZrO_2$ slab with the distance $R_Z$ ~45 Å along axis Z (see Methods). The simulations were performed with initial annealing temperature of 4000 K. The results of these simulations are shown in Figure 5 shows the initial immobilization of the 13FN3-14FN3 domain on the flat $ZrO_2$ surface with the lowest adsorption energy (config_flat, Table 1).

Table 1. The calculated adsorption energy for atomically flat and nanostructured surface

| Surface type | Immobilized configuration (notation) | Total adsorption energy (kcal/mol) | EI (kcal/mol) | vdW (kcal/mol) |
|---|---|---|---|---|
| Flat surface | config_flat | -89.3 | -0.35 | -88.95 |
| Nanostructured surface | config_pyr_1 | -66.1 | -37 | -29 |
| | config_pyr_2 | -97.4 | -67 | -30.4 |
| | config_pyr_2.1 | -31 | EX | -31 |
| | config_pyr_3 | -30.4 | EX | -30.4 |

Denotes: "EX" denotes that EI are excluded from the total adsorption energy calculations.

The Figure 5 indicates that the vdW inter-atomic interactions (between the protein and the atomically flat surface) are responsible (almost entirely) for adsorption of the rigid protein body on the substrate surface. This result is physically understood because the considered $ZrO_2$ atomic flat dielectric slab is stoichiometric and does not possess a net charge. That is why summation of the pair-wise EI between partial charged protein atoms and flat ZrO$_2$ slab atoms (with 2:1 ratio of negative and positive charged O and Zr atoms, respectively) results in negligible contribution of the electrostatic energy component to the total energy adsorption (see Table 1). Comparing the configuration of the lowest adsorption energy with the other metastable immobilized configurations (not shown), we obtain the physically expected result[27]: the larger the number of protein residues strongly interacting with the substrate (contact area), the larger the total attractive vdW inter-atom interactions and larger adsorption energy (Figure 5). It is interesting to note, that in the case (config_flat) of the lowest adsorption energy (Figure 5), the rather large part of the adsorption energy is provided by amino acid residues (Figure 5B) involved in the HBS located at the 13FN3 fragment. These residues have large, extended side chains that can provide multiple inter-atom vdW interactions.

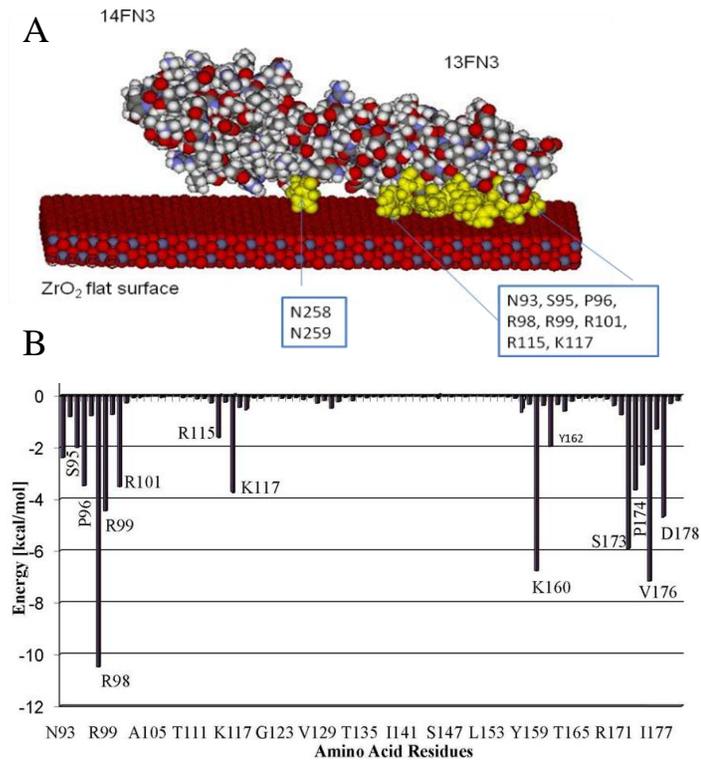

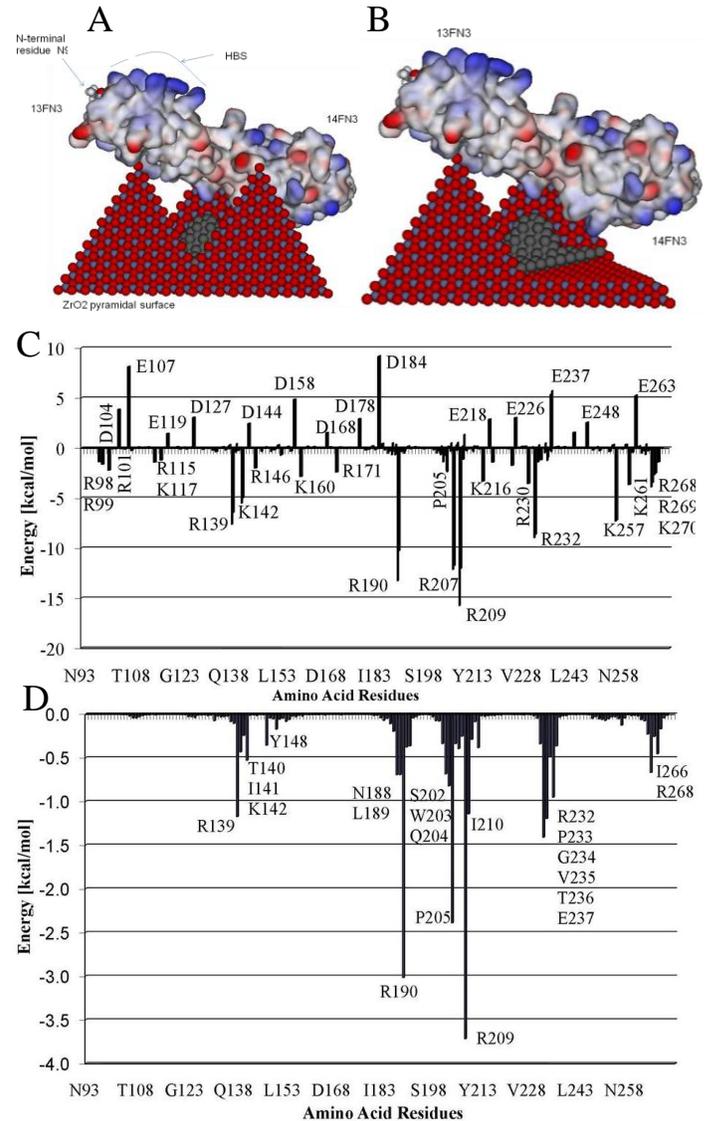

Figure 5. (**A**) Immobilization of the 13FN3-14FN3 fragment on the ZrO$_2$ flat surface taking into account the vdW and EI between the protein and surface atoms. Yellow shows the residues (detail shown in figure 5B) responsible for adsorption due to vdW interactions with the ZrO$_2$ slab. (**B**) The contribution (major) of each amino acid residue (from the N93-S180 sequence of the 13FN3 fragment) to the total adsorption energy due to vdW interactions.

To compare the initial adsorption of the protein on the flat and nanostructured surfaces, we perform MC simulations in search of the optimal immobilization of the 13FN3-14FN3 domain on the nanostructured surface of the ZrO$_2$ slab using a model pyramidal fragment (Figure 2B). We perform two MC runs (config_pyr_1): first, with an initial annealing temperature of 500K (Figure 6); and, second (config_pyr_2) - with an initial annealing temperature of 4000K (Figure 7). Both simulations are carried out with the starting position of the protein where the geometrical centers for the protein and the ZrO$_2$ slab were separated by the distance R$_z$ ~39 Å along axis Z (see Methods).

Figure 6. (**A**) Immobilization of the 13FN3-14FN3 fragment on a slab of ZrO$_2$ nanostructured pyramidal surface (config_pyr_1) shown by a colored solid solvent-accessible protein surface that represents the electrostatic potential distribution for positively charged (Arg/Lys) basic (blue), negatively charged (Asp/Glu) acidic (red) and neutral (white) amino acid residue regions. The N-terminal neutral amino acid residue N93 of the fragment and the HBS are both shown by arrow. (**B**) This figure is analogous to figure A, with one of the large pyramids removed. The central small pyramid is shown by gray. (**C**) The contribution of each amino acid residue (from the N93-T271 sequence of the 13FN3-14FN3 fragment) to the total adsorption energy (example _pyr_1) due to both the electrostatic and vdW interactions. (**D**) The contribution of each amino acid residue (from the N93-T271 sequence of the 13FN3-14FN3 fragment) to the total adsorption energy (config_pyr_1) due to vdW interactions.

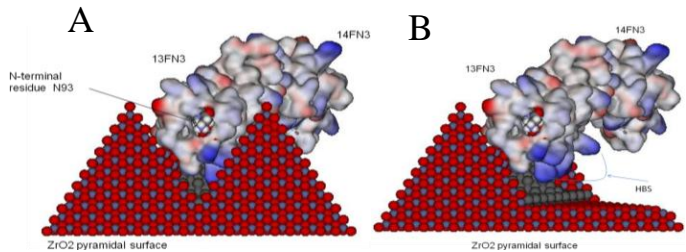

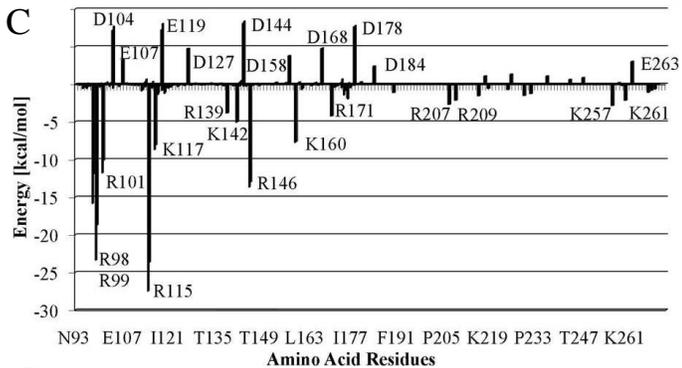

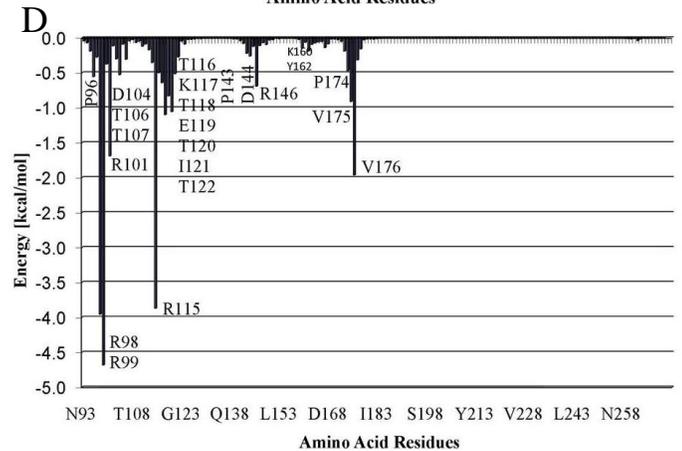

Figure 7. Immobilization of the 13FN3-14FN3 fragment on a slab of ZrO$_2$ nanostructured pyramidal surface (config_pyr_2). The figures A, B, C, D, for this case and the corresponding legends are analogous to Figure 6 (config_pyr_1).

Figures 6 and 7 show two examples of immobilized configurations of 13FN3-14FN3 domain on the Zr surface, which correspond to locally stable configurations with adsorption energies: -66.1 kcal/mol (config_pyr_1) and -97.4 kcal/mol (config_pyr_2), respectively (see Table 1). In the first configuration (config_pyr_1, Figure 6), ~44% (-29 kcal/mol) of the adsorption energy is contributed by the vdW interactions, while ~56% (-37kcal/mol) of the energy is due to EI. In the second configuration (config_pyr_2, Figure 7), the adsorption energy was found to be significantly larger due to the increase of the electrostatic component contribution, which is ~ 69% (-67kcal/mol) of the total energy, while the corresponding vdW contribution was ~31% (-30.4kcal/mol). As we can see from

Figures 6C and 7C, the initial immobilization of the protein structure is a result of attractive and repulsive EI of the basic (positive charged Arg/Lys) and acidic (negative charged Asp/Glu) residues with the negatively charged sites of nanostructured surface. It should be noted that in this process the negatively charged amino acid residues (Figure 7C) determine and facilitate the preferable orientation due to effective repulsion from negatively charged pyramidal edges and vertexes. Our calculations show that the total contribution to the adsorption energy associated with EI exceeds the contribution due to vdW interactions.

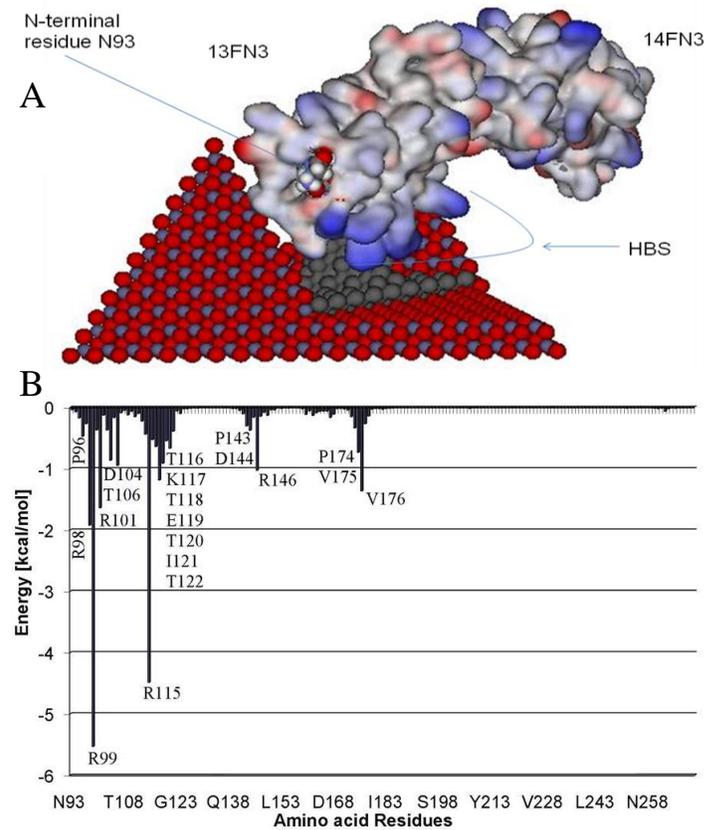

Figure 8. (**A**) Immobilization of the 13FN3-14FN3 fragment on a slab of ZrO$_2$ nanostructured pyramidal surface (started from the configuration of config_pyr_2 taking into account only vdW interactions) shown by a colored solid solvent-accessible protein surface that represents the electrostatic potential distribution for positively charged (Arg /Lys) basic (blue), negatively charged (Asp/ Glu) acidic (red) and neutral (white) amino acid residue regions. The N-terminal neutral amino acid residue N93 of the fragment and the HBS are both shown by arrows. The ZrO$_2$ pyramidal surface fragment is shown here with one of the large pyramidal hillocks removed for simplicity. (**B**) The contribution of each amino acid residue (from the N93-T271 sequence of the 13FN3-14FN3 fragment) to the total adsorption energy due to only vdW interactions.

As can be seen from Figures 6 and 7, the protein on the surface can be immobilized with orientations of the HBS at the surface (towards and away, respectively) in locally stable configurations.

The results of the MC simulations (Figure 7, A and B) and energy calculations (Table 1) show that the protein has a preferable orientation in the immobilized state on the substrate surface (config_pyr_2), which is provided by strong electrostatic attractions of the positively charged amino acid residues of the HBS (see Methods) to the two negatively charged edges of one large pyramid and the corresponding edge of the small pyramid. It should be noted that the major vdW component of the adsorption energy is also associated with the residues from the HBS (Figure 7D). This effect has a simple explanation. First, there are strong electrostatic attractions between the positively charged ionogenic groups of the basic amino acid residues (Arg/Lys) located at the end of the corresponding side chains and the oppositely charged pyramidal edges. Simultaneously, extended non-polar parts (carbon chain attached to the α–carbon) of the corresponding side chains are in the vicinity of the (111) plane surface between the above two pyramid edges, which are responsible for relatively large number of the close interatomic contacts. These contacts provide large vdW interactions (from -0.5 to -4.7 kcal/mol) between corresponding amino acid residues and the above surface, including its edges (Figure 7D).

To check the consistency of these results, we carry out MC simulations and energy calculations without EI in the model adsorption potential (1) for two starting configurations. We started MC simulations in the first case from the protein orientation obtained in immobilized state (config_pyr_2) with the minimal adsorption energy (Figure 7A). An obtained immobilized configuration (config_pyr_2.1) is shown in Figure 8. The starting position of the protein in the second simulation, where the both geometrical centers for the protein and the surface fragment (Figure 2B) were shifted apart on the given distance $R_Z$ ~39 Å along axis Z (see Methods), resulted in the sufficiently different immobilized configuration (config_pyr_3) shown in Figure 9. Both simulations are performed with an initial annealing temperature of 500K. These simulations show that protein can be immobilized in two distinct orientations relative to the surface, and that the adsorption energy is very close in both cases (Table 1). . It suggests that vdW interactions are not selective in terms of the preferable orientation of protein in the immobilization process.

Comparison between two protein immobilized configurations, config_pyr_2.1 (Figure 8) and config_pyr_2 (Figure 7), shows similar orientation of the immobilized proteins obtained with different initial annealing temperatures in MC simulations, as well as a very similar corresponding vdW component of the adsorption energies (Table 1, -31kcal/mol and -30.4kcal/mol, respectively). This similarity suggests that the EI (due to the charged HBS cluster) govern the particular orientation of the protein on the nanostructured $ZrO_2$ surface.

Analogous comparison between protein immobilization found in config_pyr_3 (Figure 9) and config_pyr_1 (Figure 6) shows some similarity in protein orientation on the surface, mainly due to "exposed" orientation of the HBS site relative to the surface, and similar vdW components of the adsorption energies (Table 1, -31 kcal/mol and -29 kcal/mol, respectively). As one can see from Figure 9B and Figure 6C, many identical amino acid residues (R139, K142, N188, L189, Q204, R190, I266 and R268) are responsible for vdW interactions in the both cases. This similarity suggests again, that the favorable EI are the driving force that determines the basic orientation of the protein on the surface. The similar contribution to the protein adsorption energy due to vdW interactions (~-30 kcal/mol) on the nanostructured surface can be achieved in multiple immobilized protein orientations. The EI contribute to both the strength of the protein adsorption and to the protein orientation with respect to the surface.

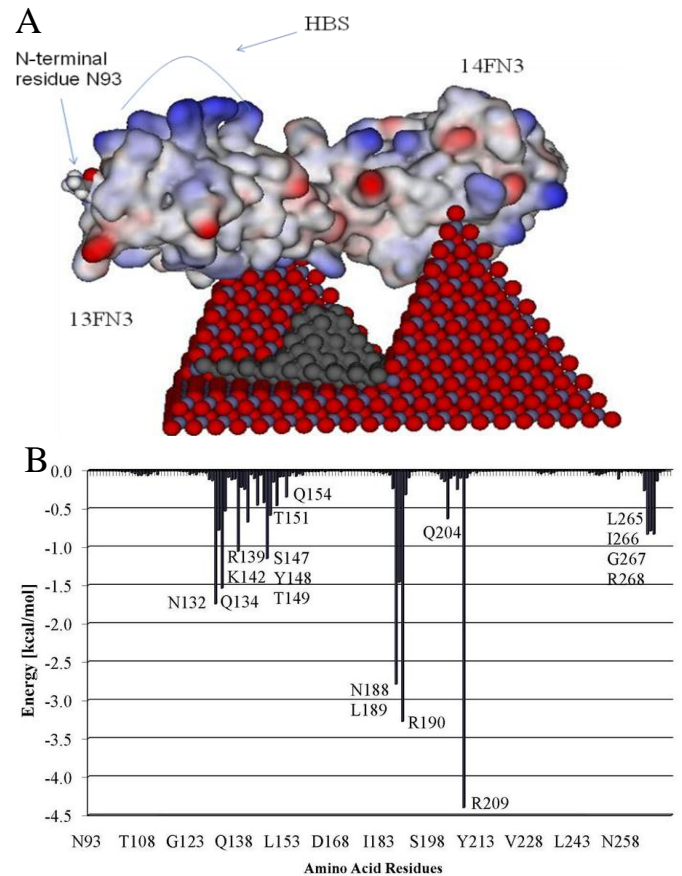

Figure 9. (**A**) Immobilization of the 13FN3-14FN3 fragment on a slab of $ZrO_2$ nanostructured pyramidal surface (config_pyr_3; started from the basic initial orientation of FN with RZ=39 A, taking into account only vdW interactions) shown by a colored solid solvent-accessible protein surface that represents the electrostatic potential distribution for positively charged (Arg/Lys) basic (blue), negatively charged (Asp/Glu) acidic (red) and neutral (white) amino acid residue regions. The N-terminal neutral amino acid residue N93 of the fragment and the HBS site are both shown by arrow. For simplicity, the $ZrO_2$ pyramidal surface fragment is shown with one of the large pyramids removed hillock. (**B**) The contribution of each amino acid residue (from the N93-T271 sequence of the 13FN3-14FN3 fragment) to the total adsorption energy due to only vdW interactions.

Overall, our MC simulations show that the optimal protein immobilization on the nanostructured pyramidal $ZrO_2$ surface

(config_pyr_2, Figure 7) has the absolute value of adsorption energy larger than what is typical for the protein adsorption on the smooth (flat) surface (config_flat, Figure 5). Being negligibly small in the case of adsorption on the smooth surface, the strong attractive EI play a major role in the increase of protein adsorption on the nanostructured surface. Our model simulations of protein adsorption on nanostructured zirconia surface demonstrate the increase of electrostatic interactions with protein compared to the atomically flat surface, while the vdW interactions are reduced. This is due to the specific selection of the surface geometry, which is a limitation of our model. Ideally, the vdW interactions should also increase at the nanostructured surface due to larger available surface contact area. One of the immediate suggestion for an increase of the adsorption energy of the protein on the nanostructured $ZrO_2$ surface is a moderate increase of the pyramidal hillocks size to create condition for the larger contact area (i.e. better steric fit) and as a result the stronger vdW interactions. The slight enlargement of pyramid size should increase the vdW contribution to the adsorption energy due to the larger surface contact area, and the negatively charged pyramidal edges should still provide the favorable EI. In this way, we may obtain conditions for the significant increase of adsorption energy due to favorable electrostatic and vdW interactions between the protein and nanostructured surface. In other words, there is a possibility to significantly increase protein adsorption on the considered surface and control the protein adsorption by changing the topology of nanostructured engineered cubic $ZrO_2$ coatings produced, (for example) by IBAD technique at different deposition conditions[23,24]. In our future work we will develop a simulated surface growth technique to maximize vdW interaction between protein and the surface. This approach will predict the inorganic surface optimal for protein adsorption with a specific orientation.

The obtained result of the protein immobilization on the inorganic solid surface points to the importance of electrostatic and steric complementarity. This effect is well-known in protein-protein interactions in the formation of known protein complexes[44,70-73]. This complementarity corresponds to electrostatic and steric fit between interacting proteins. The steric fit emphasizes the vdW component, while the electrostatic fit is a long-range electrostatic component of the adsorption energy. Thus, analogous to the problem of the protein-protein complex prediction, the best electrostatic and steric fit of the protein to the inorganic surface (associated with initial protein immobilization or adsorption) corresponds to a minimum of the adsorption energy determined by the non-covalent interactions.

## CONCLUSION

In the present work, we find the optimal immobilization of the rigid 13FN3-14FN3 fragment on the model atomic nanostructured and smooth $ZrO_2$ surfaces. We show that the protein immobilization on the nanostructured pyramidal $ZrO_2$ surface is achieved with a absolute value of adsorption energy larger than the adsorption energy of the protein on the smooth (flat) surface. The strong attractive EI, while negligible in the case of adsorption on the flat surface, play the major role in the increased adsorption strength on nanostructured Zr surface. We also show that nanostructured surface significantly modifies the orientation of the adsorbed protein relative to the surface. This observed influence on the orientation and adsorption strength of the protein can be used to promote the cell adhesion, because the corresponding cell integrin receptors are able to bind with the RGD sites of the immobilized proteins exposed into solution. We obtain this result using a multidisciplinary approach, which combines the solid state physics and computational molecular biophysics developed in this work to describe the initial protein adsorption on the solid surface. First, based on the first principle, quantum mechanical calculations for the $ZrO_2$ nanostructure in the shape of a pyramid with three-fold symmetry, we found a significant variation of the charge density and the electrostatic potential on its surface. The surface features such as edges and vertices cause this non-uniformity, while the smooth (flat) $ZrO_2$ surface does not show the variation of the charge density across its surface because of its translational symmetry. Second, we implemented the MC simulated annealing method using a Metropolis algorithm to find the optimal immobilization of the protein body on the nanostructured surface that corresponds to the minimum adsorption energy.

Our results suggest that a major physicochemical mechanism in initial protein adsorption on the non-organic solid nanostructured surface is due to the favorable long-range EI between specific surface charges (edges and vertexes) and oppositely-charged amino acid residues on the protein surface. The vdW interactions become significant at short distances between the protein and the non-organic surface and may slightly adjust the final orientation of the immobilized protein on the surface. We show that in our calculations the optimal electrostatic and steric fit of the protein to the inorganic surface corresponds to a minimum of the adsorption energy determined by the non-covalent interactions.

## ACKNOWLEDGMENT
We thank Dr. Leo Kinarsky (University Nebraska Medical Center, Omaha) for fruitful discussions. AR, RFS and FN work was supported by Nebraska Research Initiative and NSF.